\documentclass{JHEP}
\usepackage{amsmath,amssymb}
\usepackage{epsfig}
\usepackage{cite}

\newcommand{\eq}{\begin{equation}}
\newcommand{\feq}{\end{equation}}
\newcommand{\eqn}{\begin{eqnarray}}
\newcommand{\feqn}{\end{eqnarray}}
\newcommand{\arr}{\begin{eqnarray*}}
\newcommand{\farr}{\end{eqnarray*}}

\font\mybb=msbm10 at 12pt
\def\bb#1{\hbox{\mybb#1}}

\def\bR {\bb{R}}

\newcommand{\cosech}{\,{\rm cosech}\,}

\title{The Asymptotic Dynamics of two-dimensional (anti-)de~Sitter Gravity}
\author{Mauro Brigante \\
Dipartimento di Fisica dell'Universit\`a di Milano \\
Via Celoria 16, I-20133 Milano.\\
E-mail: \email{maubriga@tin.it}}
\author{Sergio Cacciatori \\
Dipartimento di Matematica dell'Universit\`a di Milano \\
Via Saldini 50,
I-20133 Milano and \\
INFN, Sezione di Milano, Via Celoria 16, I-20133 Milano.\\
E-mail: \email{cacciatori@mi.infn.it}}
\author{Dietmar Klemm and Daniela Zanon \\
Dipartimento di Fisica dell'Universit\`a di Milano \\
Via Celoria 16, I-20133 Milano and \\
INFN, Sezione di Milano, Via Celoria 16, I-20133 Milano. \\
E-mail: \email{dietmar.klemm@mi.infn.it,
               daniela.zanon@mi.infn.it}}
\preprint{IFUM-707-FT \\
hep-th/0202073}

\abstract{We show that the asymptotic dynamics of two-dimensional
de~Sitter or anti-de~Sitter Jackiw-Teitelboim (JT) gravity is described
by a generalized two-particle Calogero-Sutherland model.
This correspondence is established by formulating
the JT model of (A)dS gravity in two dimensions as a
topological gauge theory, which
reduces to a nonlinear 0+1-dimensional sigma model on the
boundary of (A)dS space. The appearance of cyclic coordinates
allows then a further reduction to the
Calogero-Sutherland quantum mechanical model.}

\keywords{AdS-CFT Correspondence, 2D Gravity, Field Theories in Lower
Dimensions}

\begin{document}

\section{Introduction}
\label{intro}

There is by now a great deal of evidence that the correspondence
between type IIB string theory on AdS$_5$ $\times$ S$^5$ and
${\cal N}=4$ super-Yang-Mills theory in four
dimensions \cite{Maldacena:1997re} extends
to a more general duality between any gravitational theory on
anti-de~Sitter spaces and conformal field theories residing on the
boundary of AdS \cite{Witten:1998qj}. Such a duality represents
an explicit realization of the holographic principle \cite{'tHooft:gx}.
In three dimensions, pure Einstein-AdS gravity was known to be
related to a two-dimensional conformal field theory \cite{Coussaert:1995zp}
even before the advent of the AdS/CFT correspondence. This is based
on the fact that pure Einstein gravity in three dimensions can be written
as a Chern-Simons theory \cite{Achucarro:1986vz}, which is known to
reduce to a WZNW model in presence of a boundary \cite{Elitzur:1989nr}.
The boundary conditions for asymptotically AdS spaces \cite{Brown:nw}
provide then the constraints for a Hamiltonian reduction from
the WZNW model to Liouville field theory \cite{Coussaert:1995zp}.
Recently the equivalence of pure gravity in three dimensions
and two-dimensional Liouville theory has been extended
in \cite{Cacciatori:2001un} to the case of positive cosmological constant,
providing thus an explicit example of the proposed correspondence between
de~Sitter gravity and Euclidean conformal field
theories \cite{Strominger:2001pn}.

Much less is known for two-dimensional anti-de~Sitter
gravity \cite{Strominger:1998yg}, which, in the spirit of AdS/CFT,
should have a dual description in terms of a conformal quantum
mechanical system \cite{Claus:1998ts}\footnote{Cf.~also
\cite{Cacciatori:1999rp} and references therein.}. Progress in this
direction has been made in \cite{Cadoni:2000gm}, where it was shown
that two-dimensional Jackiw-Teitelboim-AdS gravity \cite{JT}
induces on the spacetime boundary a conformally invariant dynamics that
can be described in terms of a de Alfaro-Fubini-Furlan
model \cite{deAlfaro:1976je} coupled to an external source.

In the present paper we will derive a correspondence between
two-dimensional AdS or dS gravity and a quantum mechanical system
in an alternative way, using the fact that the Jackiw-Teitelboim (JT)
model can be formulated as a topological gauge
theory \cite{Isler:1989hq,Chamseddine:1990wn}. It will turn out that,
in the same way in
which a Chern-Simons theory in three dimensions reduces to a WZNW
model in presence of a boundary, this 2d topological gauge theory
reduces to a nonlinear $\sigma$ model on a one-dimensional
boundary.

The remainder of this paper is organized as follows: In section
\ref{topgaugeth}, the formulation of the JT model of 2d dilaton gravity
as a topological gauge theory is briefly reviewed. In section
\ref{asbeh}, we translate the boundary conditions for asymptotically
AdS$_2$ spaces \cite{Cadoni:1998sg,Cadoni:1999ja} into conditions
for the fields appearing in the gauge theory. In the following section,
it is shown that one must add boundary terms to the bulk action
in order to have a well-defined variational principle. Since the bulk
action is topological, the dynamics of the system is entirely encoded
in these boundary terms, which give rise to a nonlinear $\sigma$ model in
$0+1$ dimensions. In section \ref{furtherred} we perform a further
reduction of the $\sigma$ model to a generalized two-particle
Calogero-Sutherland model.
This is similar to the familiar Hamiltonian reduction
from the WZNW model to Liouville theory. Finally, in section \ref{poscosm},
we generalize the considerations of sections \ref{asbeh}-\ref{furtherred},
which refer to AdS gravity, to the case of positive cosmological constant.

\section{Two-dimensional (anti-)de~Sitter gravity as a
topological gauge theory}
\label{topgaugeth}

It is well-known that the Jackiw-Teitelboim (JT) model \cite{JT}
of dilaton gravity in
two dimensions, with action given by

\begin{equation}
S = \frac 12 \int d^2x \sqrt{-g} \eta (R - 2\Lambda)\,, \label{actiongrav}
\end{equation}

where $\eta$ denotes the dilaton and $\Lambda$ the cosmological constant,
can be formulated as an SU(1,1) topological gauge
theory \cite{Isler:1989hq,Chamseddine:1990wn}.
This is similar to the three-dimensional case,
where pure Einstein gravity can be written as a Chern-Simons
theory \cite{Achucarro:1986vz}. Apart from the gauge field, the gauge
theory contains also a scalar field $\Phi$ in the adjoint of SU(1,1).
The action reads\footnote{Note that (\ref{actiongauge}) can also be obtained
by dimensional reduction of the 3d Chern-Simons action.}

\begin{equation}
S = -\int {\mathrm{Tr}} (\Phi F)\,, \label{actiongauge}
\end{equation}

where $\Phi = \Phi^A \tau_A$ and $F = \frac 12 F^A_{\mu\nu} \tau_A dx^{\mu}
\wedge dx^{\nu}$ take values in the Lie algebra su(1,1). (The generators
$\tau_A$ are given in the appendix). The field strength is given by

\begin{equation}
F_{\mu\nu} = \partial_{\mu} A_{\nu} - \partial_{\nu} A_{\mu}
+ [A_{\mu}, A_{\nu}]\,.
\end{equation}

The action (\ref{actiongauge}) is invariant under the gauge
transformations

\begin{equation}
A_{\mu} \to g^{-1} A_{\mu} g + g^{-1} \partial_{\mu} g\,,
\qquad \Phi \to g^{-1} \Phi g\,. \label{gaugetransf}
\end{equation}

The equations of motion following from (\ref{actiongauge}) read

\begin{eqnarray}
F_{\mu\nu} &=& 0\,, \label{F=0} \\
D_{\mu} \Phi &\equiv& \partial_{\mu} \Phi + [A_{\mu}, \Phi] = 0\,,
                       \label{DPhi=0}
\end{eqnarray}

so that the solutions are those of flat SU(1,1) connections in
two dimensions, i.~e.~, locally one can write
$A_{\mu} = g^{-1} \partial_{\mu} g$, with $g \in$ SU(1,1).\\
To make contact with gravity,
one decomposes the su(1,1) valued scalar and gauge fields according to

\begin{equation}
\Phi^A = (\phi^a, \eta)\,, \qquad A^A_{\mu} = (\lambda e^a_{\mu},
         \omega_{\mu})\,, \label{decomp}
\end{equation}

where $e^a_{\mu}$ $(a = 0,1)$ and $\omega_{\mu} = \omega_{01\mu}$
denote the zweibein and spin connection respectively.
$\lambda$ is related to the cosmological constant
by $\Lambda = \mp\lambda^2$, with the minus sign leading to AdS
and the plus sign to dS gravity.
The field strength $F^A_{\mu\nu}$ is written accordingly as

\begin{equation}
F^{A}_{\mu\nu} = (\lambda T^a_{\mu\nu}, {\cal R}_{\mu\nu})\,,
\end{equation}

where $T^a_{\mu\nu}$ will correspond to the torsion tensor,
and ${\cal R}_{\mu\nu}$ to the curvature
tensor \cite{Isler:1989hq,Chamseddine:1990wn}. Vanishing torsion
is enforced by the Lagrange multipliers $\phi^a$.
The equivalence (at the classical level\footnote{For a discussion of
the quantum equivalence cf.~\cite{Terao:1992pj}, where it was
shown that the theories are indeed equivalent in the light-cone
gauge. A further analysis can be found in \cite{Schaller:1993pa}.}) of
(\ref{actiongrav}) and (\ref{actiongauge}) has been shown
in \cite{Isler:1989hq,Chamseddine:1990wn}, so we refer to these papers
for the details.

\section{Asymptotic behaviour of the fields}
\label{asbeh}

We will now focus on the case of negative cosmological constant.
Due to Birkhoff's theorem of two-dimensional dilaton gravity, the general
solution to the equations of motion following from (\ref{actiongrav})
can be written in the form \cite{Christensen:1991dk,Cadoni:1994uf}

\begin{eqnarray}
ds^2 &=& -(\lambda^2 r^2 - a^2)dt^2 + (\lambda^2 r^2 - a^2)^{-1}dr^2\,, \\
\eta &=& \eta_0 \lambda r\,, \nonumber
\end{eqnarray}

where $a$ and $\eta_0$ denote integration constants.
Note that due to the presence of the dilaton, which represents a
position-dependent coupling constant, the solutions with different sign
of $a^2$ are physically inequivalent \cite{Cadoni:1994uf}. In particular,
the solution with positive $a^2$ represents a black hole with event
horizon at $r = a/\lambda$.

An asymptotically AdS$_2$ geometry \cite{Cadoni:1998sg,Cadoni:1999ja}
is one for which the metric behaves for $r \to \infty$ as

\begin{equation}
g_{tt} = -\lambda^2 r^2 + {\cal O}(1)\,, \qquad
g_{tr} = {\cal O}\left(\frac{1}{r^3}\right)\,, \qquad
g_{rr} = \frac{1}{\lambda^2 r^2} + {\cal O}\left(\frac{1}{r^4}\right)\,.
         \label{boundcondmetr}
\end{equation}

The boundary conditions (\ref{boundcondmetr}) must be completed by
giving the asymptotic behaviour of the dilaton. As discussed in detail
in \cite{Cadoni:1999ja,Cadoni:2000ah}, the appropriate boundary condition is

\begin{equation}
\eta = {\cal O}(r)\,, \label{boundconddil}
\end{equation}

with a subleading term of ${\cal O}(r^{-1})$ \cite{Cadoni:2000ah}.
When translated in terms of the fields $\Phi, A$, the boundary
conditions (\ref{boundcondmetr}), (\ref{boundconddil}) read

\begin{eqnarray}
\Phi &=& -\left[\eta_0 \lambda r(1 + \alpha(t)) + \frac{2a(t)}{r}\right] \tau_0
         + \left[\eta_0 \lambda r(1 + \beta(t)) + \frac{2b(t)}{r}\right] \tau_2
         + 2 \gamma(t) \tau_1 \,, \\
A_t &=& \left[\lambda^2 r + \frac{c(t)}{r}\right] \tau_0
        - \left[\lambda^2 r + \frac{d(t)}{r}\right] \tau_2
        + {\cal O}(r^{-2}) \tau_1 \,, \\
A_r &=& \left[r^{-1} + {\cal O}(r^{-3})\right] \tau_1 + {\cal O}(r^{-4})
        \tau_0 + {\cal O}(r^{-4}) \tau_2 \,,
\end{eqnarray}

where the functions $\alpha(t), \beta(t), \gamma(t), a(t), b(t), c(t), d(t)$,
that depend only on the boundary coordinate $t$, parametrize the subleading
behaviour\footnote{The next to leading terms of $A_r$ were not parametrized,
as they will not be used explicitely in what follows.}.

Now some comments are in order. First of all, the asymptotics of the
fields $\phi^0, \phi^1$ do not follow from (\ref{boundcondmetr}) and
(\ref{boundconddil}), but rather can be inferred from the scalar
equations of motion (\ref{DPhi=0}). Second, although one
could consider general boundary conditions allowing for $\alpha(t) \neq
\beta(t)$, we will {\it define} a configuration to be asymptotically
AdS$_2$ if $\alpha(t) = \beta(t)$. This definition is entirely consistent.

A well-defined asymptotics for the fields $\Phi$ and $A$ can be obtained
after performing a gauge transformation (\ref{gaugetransf}), with
group element $g$ given by

\begin{equation}
g = \left(\begin{array}{cc} \cosh\theta & \sinh\theta \\
                            \sinh\theta & \cosh\theta
          \end{array}\right)\,, \qquad
\cosh\theta = \frac 12\left(\sqrt{\frac fr} + \sqrt{\frac rf}\right)\,,
\label{g}
\end{equation}

where $f = f(t)$ denotes a function of $t$ to be specified later.
The transformed fields $\Phi' = g^{-1}\Phi g$ and $A' = g^{-1}A g + g^{-1}dg$
obey, to leading order, the boundary conditions

\begin{eqnarray}
\Phi' &=& -\left[\eta_0\lambda (1 + \alpha)f + \frac{a-b}{f}\right]\tau_0 
          + \left[\eta_0\lambda (1 + \alpha)f - \frac{a-b}{f}\right]\tau_2
          + 2\gamma \tau_1\,, \nonumber \\
A'_t &=& \left[\lambda^2 f + \frac{c-d}{2f}\right]\tau_0 -
         \left[\lambda^2 f - \frac{c-d}{2f}\right]\tau_2 + \frac{\dot{f}}{f}
         \tau_1\,, \label{boundcondprime} \\
A'_r &=& 0\,. \nonumber
\end{eqnarray}

For reasons that will become clear in the next section, we would like
to have the boundary condition

\begin{equation}
\Phi = \kappa(t)A_t\,, \label{boundcondfinal}
\end{equation}

with $\kappa(t)$ some function. (\ref{boundcondfinal}) can be obtained
from (\ref{boundcondprime}) by a second gauge transformation, with
group element

\begin{equation}
h = \left(\begin{array}{cc} \exp(i\varphi(t)) & 0 \\ 0 & \exp(-i\varphi(t))
          \end{array}\right)\,.
\label{h}
\end{equation}

The transformed fields $\Phi'' = h^{-1}\Phi' h$ and
$A'' = h^{-1}A' h + h^{-1}dh$ will then obey the boundary condition
(\ref{boundcondfinal}) if we choose the functions $f(t)$ and $\varphi(t)$
appearing in the transformation matrices $g$ and $h$ as
solutions of the differential equations

\begin{eqnarray}
\dot f &=& - 2\gamma f\frac{2\lambda^2 f^2 - (c-d)}{f^2\eta_0\lambda(1+\alpha)
           - (a-b)}\,, \nonumber \\
\dot{\varphi} &=& f\frac{2\lambda^2(a-b) - \eta_0\lambda(1+\alpha)(c-d)}
               {f^2\eta_0\lambda(1+\alpha) - (a-b)}\,. \nonumber
\end{eqnarray}

The function $\kappa(t)$ appearing in (\ref{boundcondfinal}) is then
given by

\begin{equation}
\kappa(t) = \frac{4f\gamma}{\dot f}\,.
\end{equation}

Note that the functions $f(t)$ and $\varphi(t)$ as well as the
proportionality "constant" $\kappa(t)$ depend on the boundary fields
$\alpha(t), \gamma(t), a(t), b(t), c(t)$ and $d(t)$. In what follows, we
will drop the primes on $\Phi''$ and $A''$.

\section{Boundary terms in the action}

In the variation of the action (\ref{actiongauge}) one encounters
a term

\begin{equation}
\int d^2x {\mathrm{Tr}}(\Phi\partial_r\delta A_t) = -\int d^2x {\mathrm{Tr}}
(\partial_r \Phi \delta A_t) + \int dt {\mathrm{Tr}}(\Phi \delta A_t)\,.
\end{equation}

This means that the action is not an extremum on-shell, instead $\delta S$
equals the surface term\footnote{Here we used the boundary condition
(\ref{boundcondfinal}).} $\int dt {\mathrm{Tr}}(\kappa(t) A_t \delta A_t)$.
In order to cancel this, we must add a boundary term to the action
(\ref{actiongauge}), leading to the improved action

\begin{equation}
S = \int d^2x {\mathrm{Tr}}(\Phi F_{rt}) - \frac 12 \int dt \kappa(t)
    {\mathrm{Tr}} \left[(A_t)^2\right]\,.
\end{equation}

The function $-\kappa(t)/2$ can be absorbed into the time coordinate by
defining $d\tau = -2dt/\kappa(t)$.

We recognize the $\Phi^A$ as Lagrange multipliers implementing the
Gauss law constraint $F^A_{r\tau} = 0$. In the functional integral
one can integrate over $\Phi$ to obtain $\delta(F_{r\tau})$,

\begin{equation}
\int DA D\Phi\, e^{iS}\ldots = \int DA \delta(F_{r\tau}) e^{i\int d\tau
                             {\mathrm{Tr}}[(A_{\tau})^2]}\ldots\,.
\end{equation}

The constraint $F_{r\tau} = 0$ is easily solved by
$A = g^{-1}dg$\footnote{As we consider the universal covering space of
AdS$_2$, holonomies can be dropped.}.
Like in \cite{Elitzur:1989nr}, we now change variables from $A$ to $g$
in the path integral,

\begin{equation}
\int DA \delta(F_{r\tau}) e^{i\int d\tau {\mathrm{Tr}}[(A_{\tau})^2]}
\ldots = \int Dg\, e^{i\int d\tau {\mathrm{Tr}}
[(g^{-1}\partial_{\tau}g)^2]}\ldots\,.
\end{equation}

We have thus derived an effective action

\begin{equation}
S_{{\mathrm{eff}}} = \int d\tau {\mathrm{Tr}}
\left[(g^{-1}\partial_{\tau}g)^2\right] \label{actsigma}
\end{equation}

for the problem under consideration. (\ref{actsigma}) describes a
nonlinear $\sigma$ model in 0+1 dimensions.

If we write $g \in$ SU(1,1) in the Iwasawa decomposition

\begin{equation}
g = \left(\begin{array}{cc} e^{iu/2} & 0 \\ 0 & e^{-iu/2} \end{array}\right)
    \left(\begin{array}{cc} \cosh\frac{\vartheta}{2} &
    \sinh\frac{\vartheta}{2} \\ & \\
    \sinh\frac{\vartheta}{2} & \cosh\frac{\vartheta}{2} \end{array}\right)
    \left(\begin{array}{cc} e^{iv/2} & 0 \\ 0 & e^{-iv/2} \end{array}\right)\,,
\end{equation}

the action (\ref{actsigma}) reads

\begin{equation}
S_{{\mathrm{eff}}} = \frac 12 \int d\tau (\dot{\vartheta}^2 - \dot u^2 -
                     \dot v^2 - 2\dot u \dot v \cosh \vartheta)\,,
                     \label{actparam}
\end{equation}

where a dot denotes differentiation with respect to $\tau$.
Defining $x^1 = u$, $x^2 = v$, $x^3 = \vartheta$, one can write

\begin{equation}
S_{{\mathrm{eff}}} = \frac 12 \int g_{ij}\dot x^i \dot x^j d\tau\,,
\end{equation}

where $i,j = 1,2,3$ and

\begin{equation}
(g_{ij}) = \left(\begin{array}{ccc} -1 & - \cosh \vartheta & 0 \\
                                  - \cosh \vartheta & -1 & 0 \\
                                  0 & 0 & 1 \end{array} \right)
\label{metrgrp}
\end{equation}

is a constant curvature metric on
the group manifold SU(1,1). (\ref{metrgrp}) admits the six
Killing vectors

\begin{eqnarray}
L_0 &=& - i\partial_v\,, \qquad
L_{\pm} = e^{\mp iv}[i\partial_{\theta} \pm \coth \vartheta\,
          \partial_v \mp \cosech \vartheta\, \partial_u]\,, \nonumber \\
\tilde L_0 &=& - i\partial_u\,, \qquad
\tilde L_{\pm} = e^{\mp iu}[i\partial_{\theta} \pm \coth \vartheta\,
          \partial_u \mp \cosech \vartheta\, \partial_v]\,,
\label{invaralg}
\end{eqnarray}

which satisfy the commutation relations

\begin{eqnarray}
[L_{\pm}, L_0] &=& \pm L_{\pm}\,, \qquad [L_+, L_-] = 2L_0\,, \nonumber \\
{[}\tilde L_{\pm}, \tilde L_0] &=& \pm \tilde L_{\pm}\,, \qquad [\tilde L_+,
\tilde L_-] = 2\tilde L_0\,,
\end{eqnarray}

and $[L, \tilde L] = 0$. The Killing vectors generate thus the algebra
su(1,1) $\times$ su(1,1), corresponding to the invariance of the
action (\ref{actsigma}) under $g \to h g \tilde h$, with $h, \tilde h
\in$ SU(1,1) independent of $\tau$.

\section{Further reduction}
\label{furtherred}

In this section we will show that, due to the presence of cyclic
coordinates, the $\sigma$ model (\ref{actsigma}) can be further
reduced to a simple quantum mechanical system.

As the Lagrangian in (\ref{actparam}) does not depend on $u,v$,
the conjugate momenta

\begin{equation}
\pi_u = -\dot u - \dot v \cosh\vartheta\,, \qquad
\pi_v = -\dot v - \dot u \cosh\vartheta
\end{equation}

are constants of motion. In order to implement the constraints
$\pi_{u,v} = {\mathrm{const}}$, one has to go from (\ref{actparam}) to
the reduced action (Routhian function)

\begin{equation}
S_{{\mathrm{eff}}} \to S_{{\mathrm{eff}}} - \int d\tau [\dot u \pi_u
+ \dot v \pi_v]\,,
\end{equation}

i.~e.~, one has to perform a partial Legendre transformation with respect to
$\dot u$, $\dot v$. In this way, one finally obtains

\begin{equation}
S_{{\mathrm{eff}}} = \frac 12 \int d\tau \left[\dot{\vartheta}^2 -
                     \frac{(\pi_u - \pi_v)^2}{4\sinh^2\frac{\vartheta}{2}} +
                     \frac{(\pi_u + \pi_v)^2}
                     {4\cosh^2\frac{\vartheta}{2}}\right]\,, \label{actionqm}
\end{equation}

which is a generalization of the two-particle Calogero-Sutherland
model \cite{Sutherland:1971kq}.
The relationship between AdS$_2$ gravity and the Calogero model
was first pointed out in \cite{Gibbons:1998fa}.

Note that the only generators of the conformal algebra (\ref{invaralg}) that
commute with $\pi_v = L_0$ and $\pi_u = \tilde L_0$ are $L_0$ and
$\tilde L_0$, which means that the conformal symmetry is broken by
the reduction to (\ref{actionqm}). Conformal transformations actually
relate different $\pi_{u,v}$, and therefore fixing these momenta breaks
the invariance under su(1,1) $\times$ su(1,1).

\section{Positive cosmological constant}
\label{poscosm}

In the case of positive cosmological constant $\Lambda = \lambda^2$,
the general solution to the equations of motion following from
(\ref{actiongrav}) can be written as

\begin{eqnarray}
ds^2 &=& -(a^2 - \lambda^2 r^2)dt^2 + (a^2 - \lambda^2 r^2)^{-1}dr^2\,, \\
\eta &=& \eta_0 \lambda r\,. \nonumber
\end{eqnarray}

For a more detailed discussion of the geometry of two-dimensional
de~Sitter space we refer to the appendix. Note in particular that
for $r > a/\lambda$, $r$ becomes a timelike coordinate\footnote{Of course
$\partial_r$ is always timelike for $a^2 \le 0$.},
so that the dilaton is now time-dependent outside the static patch.

In what follows, we will be particularly interested in the behaviour
of the gravitational fields near the past boundary ${\cal I}^-$
at $r \to \infty$.
The boundary conditions for an asymptotically dS$_2$ geometry
can be found by analytically continuing (\ref{boundcondmetr}) and
(\ref{boundconddil}). This yields

\begin{equation}
g_{tt} = \lambda^2 r^2 + {\cal O}(1)\,, \qquad
g_{tr} = {\cal O}\left(\frac{1}{r^3}\right)\,, \qquad
g_{rr} = -\frac{1}{\lambda^2 r^2} + {\cal O}\left(\frac{1}{r^4}\right)\,,
         \label{boundconddS}
\end{equation}

and $\eta = {\cal O}(r)$. Alternatively, one can obtain these boundary
conditions by requiring the subleading terms to fall off fast enough to
ensure that conserved charges are well-defined, but slow enough to
make the theory non-trivial. This singles out the behaviour
(\ref{boundconddS}).

At this point, one can proceed analogously to the AdS case, and translate
(\ref{boundconddS}) into boundary conditions for the gauge field
$A$ and the scalar $\Phi$. Again, these conditions turn out to be
gauge equivalent to $\Phi = \kappa(t)A_t$, $A_r = 0$, with some function
$\kappa(t)$, as in section \ref{asbeh}. Taking into account the
boundary terms\footnote{Note that dS$_2$ has two boundaries, the past
boundary ${\cal I}^-$ and the future boundary ${\cal I}^+$, so in principle
one has to consider two boundary terms, which should lead to a
quantum mechanical system living on a disconnected manifold. However,
since we are only interested in the asymptotic dynamics of the
gravitational fields near ${\cal I}^-$, we ignore surface terms
arising at ${\cal I}^+$. Of course the same problem appears in the
AdS$_2$ case, since two-dimensional anti-de~Sitter space, unlike its
higher-dimensional cousins, has two boundaries. A possible way
to eliminate the second boundary is to cut the spacetime at $r=0$,
in order that the position-dependent coupling constant represented
by the dilaton is positive \cite{Cadoni:1999ja}.}
necessary to have a well-defined variational principle,
integrating over the Lagrange multiplier $\Phi$ in the path integral,
and changing variables from $A = g^{-1}dg$ to $g$, one derives also in
the dS case an effective $\sigma$ model action (\ref{actsigma}),
which allows a further reduction to the quantum
mechanics (\ref{actionqm})\footnote{A similar reduction to a quantum
mechanical system was performed in \cite{Gegenberg:1996ye}. There
the boundary was the horizon, whereas we consider the boundaries
at infinity.}.
Note that these one-dimensional models live now in Euclidean space,
since the boundaries of dS$_2$ are spacelike.

\section{Conclusions}

To sum up, we showed that the asymptotic dynamics of two-dimensional
de~Sitter or anti-de~Sitter Jackiw-Teitelboim gravity is encoded in a
generalized two-particle Calogero-Sutherland quantum mechanical model.
We started by formulating the JT model of (A)dS gravity in two dimensions
as a topological gauge theory, which, by virtue of the (A)dS boundary
conditions, reduces to a nonlinear sigma
model on the one-dimensional boundary of (A)dS space. The appearance
of cyclic coordinates allows then a further reduction to the
quantum mechanical system (\ref{actionqm}). We have thus generalized
the known results in three dimensions, where the asymptotic dynamics
of pure Einstein gravity with negative \cite{Coussaert:1995zp}
or positive \cite{Cacciatori:2001un} cosmological constant
is described by Liouville field theory, obtained as the Hamiltonian
reduction of a WZNW model.

Note that the JT model receives no local quantum corrections at all
orders of perturbation theory \cite{Kummer:1996hy}. Probably, this
means that our results can be extended far beyond the approximation they were
actually derived.

It would also be interesting to consider the supersymmetric generalization
of the topological gauge theory (\ref{actiongauge}) \cite{Chamseddine:1990wn},
which is (at least classically) equivalent to an (A)dS supergravity
model in two dimensions. These two-dimensional supergravity theories should
then have an equivalent description in terms of a superextension of the
quantum mechanical system (\ref{actionqm}).

As also the CGHS model \cite{Callan:rs} of two-dimensional dilaton gravity
can be formulated as a topological gauge theory based on the
centrally extended Poincar\'{e}
group \cite{Cangemi:1992bj}, an analogous reduction
to a one-dimensional quantum mechanics should be possible.
The field equations following from the CGHS model in the form

\begin{equation}
S = \frac 12 \int d^2x \sqrt{-g}(\eta R - \Lambda)
\end{equation}

imply $R = 0$, so the boundary conditions are now those of asymptotically
flat metrics. One thus expects the quantum mechanical model
to be defined on a lightlike line. It would be tempting to see
explicitely what this model is.

These issues are currently under investigation.

\acknowledgments

This work was partially supported by INFN, MURST and
by the European Commission RTN program
HPRN-CT-2000-00131, in which S.~C.~, D.~K.~and D.~Z.~are
associated to the University of Torino. The authors would like
to thank R.~Jackiw, T.~Strobl and D.~Vassilevich for useful comments.
\normalsize

\newpage

\begin{appendix}

\section{Conventions}

\subsection{Anti-de~Sitter}
\label{convAdS}

An element $M$ of the Lie algebra su(1,1) satisfies

\begin{equation}
M^a_{\,\,\,b} = - \eta_{bc}\bar{M}^c_{\,\,\,d}\eta^{da}\,, \qquad
{\mathrm{Tr}} M = 0\,,\label{su11}
\end{equation}

where a bar denotes complex conjugation, and
$(\eta_{ab}) = {\mathrm{diag}}(-1,1)$. We choose as su(1,1) generators

\begin{equation}
\tau_0 = \frac 12\left(\begin{array}{cc} i & 0 \\ 0 & -i \end{array}
                 \right)\,, \quad
\tau_1 = \frac 12\left(\begin{array}{cc} 0 & 1 \\ 1 & 0 \end{array}
                 \right)\,, \quad
\tau_2 = \frac 12\left(\begin{array}{cc} 0 & -i \\ i & 0 \end{array}
                 \right)\,. \label{generatorsAdS}
\end{equation}

They are normalized according to

\begin{equation}
{\mathrm{Tr}}(\tau_A\tau_B) = \frac 12 \eta_{AB}\,,
\end{equation}

where $(\eta_{AB}) = {\mathrm{diag}}(-1,1,1)$ is the inner product
on the Lie algebra. The generators (\ref{generatorsAdS}) satisfy the relation
(\ref{su11}), and $[\tau_A, \tau_B] = -\epsilon_{ABC}\tau^C$, with
$\epsilon_{012} = 1$.

Finally, $dt \wedge dr$ is chosen to have positive orientation.

\subsection{De~Sitter}

In order to get a positive cosmological constant we interchange
$\tau_0$ and $\tau_1$ in (\ref{generatorsAdS}), i.~e.~, we take

\begin{equation}
\tau_0 = \frac 12\left(\begin{array}{cc} 0 & 1 \\ 1 & 0 \end{array}
                 \right)\,, \quad
\tau_1 = \frac 12\left(\begin{array}{cc} i & 0 \\ 0 & -i \end{array}
                 \right)\,, \quad
\tau_2 = \frac 12\left(\begin{array}{cc} 0 & -i \\ i & 0 \end{array}
                 \right)\,, \label{generatorsdS}
\end{equation}

with the normalization

\begin{equation}
{\mathrm{Tr}}(\tau_A\tau_B) = -\frac 12 \eta_{AB}\,,
\end{equation}

where now $(\eta_{AB}) = {\mathrm{diag}}(-1,1,-1)$. All the other
conventions are as in subsection \ref{convAdS}.

\section{dS$_2$ Geometry}

Consider $\bR_1^2$ with coordinates $X^0, X^1, X^2$ and metric
$(-1,1,1)$. Two-dimensional de~Sitter space can then be defined as
the hypersurface

\begin{equation}
-(X^0)^2 + (X^1)^2 + (X^2)^2 = \lambda^{-2}\,. \label{hyper}
\end{equation}

Fix now

\begin{equation}
(X^0)^2 - (X^1)^2 = \frac{r^2}{a^2} - \lambda^{-2}\,,
\end{equation}

so that $(X^2)^2 = r^2/a^2$, and parametrize the hyperbolae of fixed
$r$ by

\begin{equation}
X^0 = \frac{1}{\lambda a}\sqrt{\lambda^2 r^2 - a^2}\cosh a\lambda t\,,
\qquad X^1 = \frac{1}{\lambda a}\sqrt{\lambda^2 r^2 - a^2}\sinh a\lambda t\,.
\end{equation}

This leads to the induced metric

\begin{equation}
ds^2 = -(a^2 - \lambda^2 r^2) dt^2 + (a^2 - \lambda^2 r^2)^{-1} dr^2\,.
\end{equation}

Note that the coordinate $t$ is noncompact.

Alternatively, we can fix

\begin{equation}
(X^1)^2 + (X^2)^2 = \frac{r^2}{a^2} + \lambda^{-2}\,,
\end{equation}

so that $(X^0)^2 = r^2/a^2$, and parametrize the circles of fixed
$r$ by

\begin{equation}
X^1 = \frac{1}{\lambda a}\sqrt{\lambda^2 r^2 + a^2}\cos a\lambda t\,,
\qquad X^2 = \frac{1}{\lambda a}\sqrt{\lambda^2 r^2 + a^2}\sin a\lambda t\,.
\end{equation}

The induced metric on the hypersurface (\ref{hyper}) is now

\begin{equation}
ds^2 = -(a^2 + \lambda^2 r^2) dt^2 + (a^2 + \lambda^2 r^2)^{-1} dr^2\,.
       \label{metrdS2-}
\end{equation}

The coordinate $t$ is now compact, $t \sim t + 2\pi/\lambda a$. The
boundaries at $r \to \pm\infty$ are thus circles. $t,r$ in (\ref{metrdS2-})
are global coordinates. Defining

\begin{equation}
r = -\frac{a}{\lambda}\sinh\tau\,, \qquad t = \frac{1}{\lambda a}\phi\,,
\end{equation}

where $\phi$ is identified modulo $2 \pi$, (\ref{metrdS2-}) can be
rewritten as

\begin{equation}
\lambda^2 ds^2 = -d\tau^2 + \cosh^2\tau d\phi^2\,, 
\end{equation}

which describes dS$_2$ as a contracting/expanding one-sphere.
$\tau \to -\infty$ corresponds to the past boundary
${\cal I}^-$, whereas $\tau \to \infty$ corresponds to the future
boundary ${\cal I}^+$.

\end{appendix}

\end{document}